\documentclass[prlb,aps,amssym,nofootinbib,floatfix,twocolumn,notitlepage]{revtex4-1} 

 \usepackage{xcolor}
\usepackage{amsmath,amssymb,bbold,bm}
\usepackage{graphicx}
\usepackage{caption}
\usepackage{subcaption}
\usepackage{cancel}
\usepackage{comment}

\newcommand{\be}{\begin{equation}}
\newcommand{\ben}{\begin{equation*}}
\newcommand{\ee}{\end{equation}}
\newcommand{\een}{\end{equation*}}
\newcommand{\bs}{\begin{split}}
\newcommand{\es}{\end{split}}
\newcommand{\bmx}{\begin{array}}
\newcommand{\emx}{\end{array}}
\newcommand{\bea}{\begin{eqnarray}}
\newcommand{\bean}{\begin{eqnarray*}}
\newcommand{\eea}{\end{eqnarray}}
\newcommand{\eean}{\end{eqnarray*}}
\newcommand{\dg}{^{\dagger}}
\newcommand{\dn}{^{\vphantom{\dagger}}}

\newcommand{\lr}{\leftrightarrow}
\newcommand{\ra}{\rightarrow}

\newcommand{\ua}{\uparrow}
\newcommand{\da}{\downarrow}

\newcommand{\so}{\qquad\rightarrow\qquad}
\newcommand{\So}{\qquad\Rightarrow\qquad}

\newcommand{\andd}{\qquad\text{and}\qquad}

\newcommand{\eps}{\epsilon}

\newcommand{\sgn}[1]{{\rm sign}{#1}}
\newcommand{\pref}[1]{(\ref{#1})}

\newcommand{\intinf}[1]{\int_{-\infty}^{+\infty}{#1}}
\newcommand{\intoinf}[1]{\int_{0}^{\infty}{#1}}

\newcommand{\intob}[1]{\int_{0}^{\beta}{#1}}

\newcommand{\re}[1]{{\rm Re}\left[ #1 \right]}
\newcommand{\im}[1]{{\rm Im}\left[ #1 \right]}
\newcommand{\tr}[1]{{\rm Tr}\Big[ #1 \Big]}

\newcommand{\abs}[1]{\left\vert #1 \right\vert}
\newcommand{\bra}[1]{\left\langle #1 \right\vert}
\newcommand{\ket}[1]{\left\vert #1\right\rangle}
\newcommand{\braket}[1]{\left\langle #1\right\rangle}

\newcommand{\com}[2]{\left[#1,#2\right]}

\newcommand{\mat}[1]{\left(\bmx{cc}#1\emx\right)}

\newcommand{\sepline}{\begin{center}\rule{8cm}{.5pt}\end{center}}

\newcommand{\bw}[1]{\begin{widetext}}
\newcommand{\ew}[1]{\end{widetext}}

\newcommand{\red}[1]{{\color{red} #1}}

\begin{document}
\title{Orbital Kondo effect in fractional quantum Hall systems}
\author{Yashar Komijani$^{1,2}$, Pascal Simon$^3$ and Ian Affleck$^1$}
 \affiliation{ $^1$Department of Physics and Astronomy and $^2$Quantum Materials Institute,\\
University of British Columbia, Vancouver, B.C., Canada, V6T 1Z1 \\
$^3$Laboratoire de Physique des Solides, CNRS UMR-8502, Universit\'e Paris Sud, 91405 Orsay cedex, France
}
\date{\today}
\begin{abstract}
We study transport properties of a charge qubit coupling two chiral Luttinger liquids,  realized by two antidots placed between the edges of an integer $\nu=1$ or fractional $\nu=1/3$ quantum Hall bar. We show that in the limit of a large capacitive coupling between the antidots, their quasiparticle occupancy behaves as a pseudo-spin corresponding to an orbital Kondo impurity coupled to a chiral Luttinger liquid, while the inter antidot tunnelling acts as an impurity magnetic field. The latter tends to destabilize the Kondo fixed point for the $\nu=1/3$ fractional Hall state, producing an effective inter-edge tunnelling.  We relate the inter-edge  conductance to the  susceptibility of the Kondo impurity and calculate it analytically in various limits for both $\nu=1$ and $\nu=1/3$.	
\end{abstract}
\maketitle
\emph{{Introduction} -} 
Fractional quantum Hall (FQH) systems \cite{Yoshioka} are strongly correlated topological states, realized in clean two-dimensional electron gases under a large perpendicular magnetic field, where the bulk contains an incompressible fluid and the low energy dynamics is controlled by chiral Luttinger liquids at the edges \cite{Wen}.
There has recently been a renewed interest in these systems due to a promise of the celebrated topological quantum computation using non-abelian anyons 
\cite{Nayak08,Fiete07,Fiete0810,Ilan11} and also in connection to impurities in helical liquids of the quantum spin Hall systems \cite{Maciejko09,Tanaka11,Posske13,Vayrynen13,Vayrynen14}. 
However, there is still a considerable gap between theoretical and experimental studies of abelian anyons in the FQH edge states which motivates a more thorough study of their properties.
Here, we study the problem of elastic co-tunnelling of Laughlin quasiparticles through two antidots and show that in certain limits, it maps to a 
Kondo impurity \cite{Kondo,Hewson} embedded between two chiral Luttinger liquids  \cite{Lee92,Ingersent95,Furusaki94,Frojdh95, Kim01, Furusaki04} and exhibits interesting transport signatures. 

Transport through antidots in the FQH regime has been studied in the past, experimentally \cite{Maastilta97,Goldman0501,Arai05,Kataoka99,Sim03,Kou12} and theoretically \cite{Chamon93,Geller9700,Merlo060708,Averin07}
in a regime where the transport was dominated by correlated but incoherent transfers of individual quasiparticles. In contrast, in this paper we are interested in a regime where this sequential tunnelling is blocked due to a large inter-antidot capacitive coupling. 

Combining the pseudo-spin of a double dot with the intrinsic spin, Borda et al. \cite{Borda03} predicted an SU(4) Kondo effect which has been recently observed \cite{Keller13}. The use of double dots to realize pseudo-spin SU(2) Kondo model and its generalizations at $\nu=1$ integer quantum Hall regime was proposed in \cite{Carmi11}.
Here, we extend those ideas by studying the realization and transport properties of a Kondo impurity coupled to chiral Luttinger liquid edge states in the FQH regime. A similar model arises in the study of a double dot inserted in a spinless non-chiral Luttinger liquid, once the occupancy of the dots is limited so that they act as an effective spin. In contrast to most  previous studies that focus on zero temperature, we provide analytical expressions for the conductance in all asymptotic  temperature regimes. 
In this paper, we only deal with fully polarized (or spinless) systems and spin refers to the orbital pseduo-spin. 
\begin{figure}[h!]
\includegraphics[width=.8\linewidth]{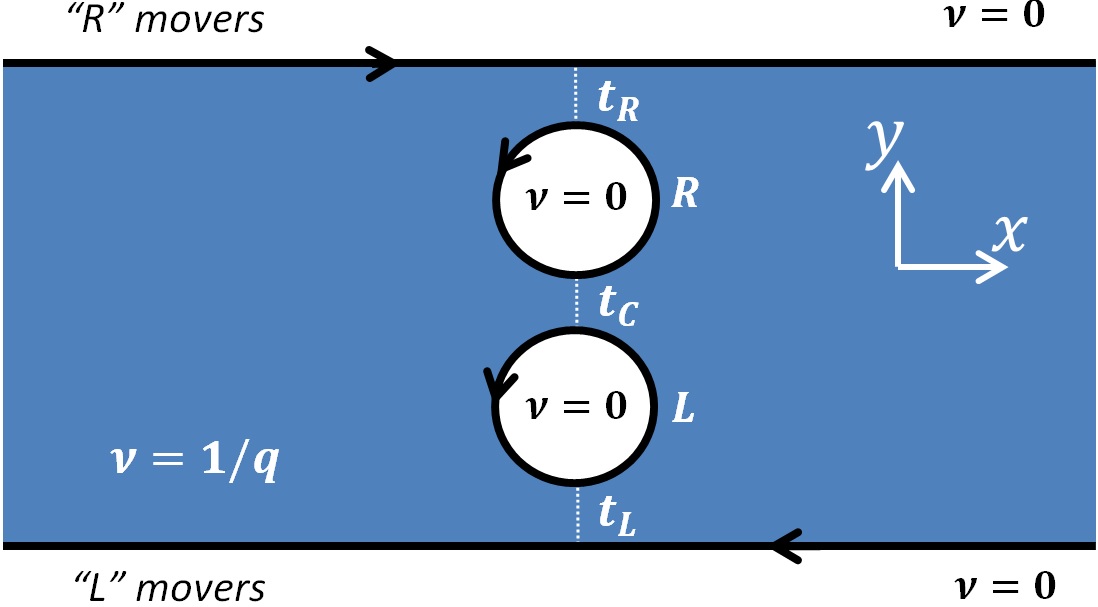}
\caption{\raggedright\small (color online) The system considered here. Two antidots enable tunnelling of quasiparticles between the outer edge states in the $\nu=1/q$ Laughlin FQH liquids. The gapped incompressible liquid (blue) plays the role of tunnel barrier for the quasiparticles. Only one state per antidot is considered. The capacitive coupling between the antidots is large enough to keep their total relative occupancy constant.}\label{fig:scheme}
\end{figure}


\emph{{The model} -}
We consider the system depicted in Fig.\,\ref{fig:scheme}, in which each antidot is represented by a single fermionic quasiparticle level. This is valid for small enough antidot  radius. In this limit the sytem can be described by the following Hamiltonian
\bea H&=&H_0-[t_R\psi\dg_{qp,R}(0)d_R+t_L\psi\dg_{qp,L}(0)d_L+h.c.]\nonumber \\&&
+U(d\dg_Rd\dn_R+d\dg_Ld\dn_L-1)^2-t_C(d\dg_Rd\dn_L+h.c.),\eea
where $t_L,t_R,t_C$ are tunnelling amplitudes and $U$ is the Coulomb energy.
Here $d_{L/R}$ annihilates quasiparticles on the upper/lower antidot and $\psi_{qp,R/L}(x)$ annihilates right/left-moving quasiparticles on the upper/lower edge of the 
Hall bar, with corresponding Hamiltonian $H_0$. 
We are interested in a parameter regime $T,t_L,t_R,t_C\ll D$, where $D\sim\min(\delta\eps,U)\ll \Delta$. Here $\delta\eps$ is the antidot level-spacing and 
$\Delta$ is the bulk energy gap (we set $k_B=1$ throughout the paper).  Then $U$  limits the antidots charge configuration to $(0,1)$ and $(1,0)$ sectors.  Sequential tunnelling is blocked in this large $U$ limit and different methods 
must be developed to study the system. 
Temporarily ignoring the inter-dot tunnelling, $t_C$, we see that to transfer one quasi-particle from upper to lower edge, we must
start in a state with the lower dot occupied, pass through a high energy intermediate state with both dots occupied or empty and end up with only 
the upper dot occupied.  Thus it is convenient to identify $L$ and $R$ with pseudo-spin up and down respectively. A Schrieffer-Wolff transformation \cite{SchriefferWolf}
then yields a  Kondo model with impurity pseudospin operators $\vec S\equiv d^\dagger_\alpha\vec \sigma_{\alpha \beta}d_\beta /2$ and 
quasiparticle pseudospin density $\vec {\cal J} (x)\equiv \psi^\dagger_{qp,\alpha}(x)\vec \sigma_{\alpha \beta}\psi_{qp,\beta}(x)/2$. The Kondo interaction, 
$J_\perp [S^x{\cal J}_x(0)+S^y{\cal J}_y(0)]+J_zS^z{\cal J}_z(0)$ contains Kondo couplings $J_\perp = 4t_Lt_R/U$, $J_z=2(t_L^2+t_R^2)/U+\delta J_z$, 
where $\delta J_z$ is an additional positive contribution arising from the Coulomb interaction between quasiparticles on the antidots and edges \cite{Averin07,fnote1}.
 Inter-antidot tunnelling corresponds to a magnetic field term in the Kondo Hamiltonian, coupled to the impurity spin only, $-t_CS^x$.

While $\psi_{qp,L/R}(x)$ are simply  free chiral fermion fields for the integer Hall state occurring at $\nu =1$; for $\nu =1/3$ it is very useful to bosonize, 
$\psi_{qp,R/L}(x)\propto e^{\pm i\varphi_{R/L}(x)}$
in terms of chiral bosons $\varphi_{R/L}(x)$, obeying the basic commutation relation $[\varphi_{R/L}(x),\varphi_{R/L}(y)]=\pm i\pi\nu\sgn(x-y)$.  Then $H_0=[v/(2\pi \nu )]\int_{-\infty}^\infty dx[(\partial_x\varphi_R)^2
+(\partial_x\varphi_L)^2]$, where $v$ is the quasiparticle velocity.
It is then convenient to define commuting right-moving spin and charge bosons, 
$\varphi_{s,c}(x)\equiv[\varphi_R(x)\pm\varphi_L(-x)]/\sqrt{2}$ since only the spin boson appears in the Kondo interaction. Then we obtain 
${\cal J}_-\equiv{\cal J}_x-i{\cal J}_y\propto e^{i\sqrt{2}\varphi_s}$ and 
$ {\cal J}_z= \partial_x\varphi_s/(2\pi \nu\sqrt{2})$.  The renormalized Kondo couplings grow larger as the 
energy scale is reduced \cite{PKA}, becoming large at the crossover scale $T_K$. For the $\nu =1$ case, $T_K=De^{-1/\lambda}$ where $\lambda \equiv \varrho J$, 
$\varrho$ is the density of states and we have assumed $J_\perp=J_z$ (similar behaviour occurs in the anisotropic case). For $\nu =1/3$, $e^{i\sqrt{2}\varphi_s}$ has 
renormalization group (RG) scaling dimension 1/3. Thus $\lambda_\perp (E)=(D/E)^{2/3}\lambda_\perp$, so $T_K\propto D/\lambda_\perp^{3/2}$.

\emph{{Conductance} -} We are interested in the inter-edge tunnelling  conductance, corresponding to back-scattering, 
defined using the charge
current operator $I=i\nu et_C(d\dg_Ld\dn_R-h.c)=2\nu et_CS^y$.
In the linear response regime  the Kubo formula gives \cite{Borda03}
\be
G=-{8\pi\nu t_C^2}\lim_{\omega\ra 0}\frac{\chi''_{yy}(\omega)}{\omega},\label{eqcon}
\ee
in units of $\nu e^2/h$ where $\chi''_{yy}$ is the imaginary part of the dynamical impurity spin susceptibility of the Kondo model, 
$\chi_{yy}(\omega )\equiv -i\int_0^\infty dt e^{i\omega t}\langle [S^y(t),S^y(0)]\rangle$. 
 Every transmitted quasiparticle contributing to the transport involves a spin-flip process at the impurity, relating conductance to the spin relaxation.

\emph{{High temperatures: $T_K,t_C\ll T$} - }
We may attempt to calculate the susceptibility using perturbation theory, but this gives a result which diverges at $\omega \to 0$:
\be 
\chi''_{yy}(\omega)=-\frac{\pi\nu}{2\omega}[\gamma^{(\nu)}_z\lambda_z^2+\gamma^{(\nu)}_\perp\lambda_\perp^2 (T/D)^{2\nu-2}]\label{eqchip}
\ee
Here $\gamma^{(\nu)}_z$ and $\gamma^{(\nu)}_\perp$ are dimensionless coefficients of ${\cal O}(1)$. 
This surprising infrared divergence is not connected to the usual renormalization of the Kondo couplings since $T\gg T_K$. Nonetheless, it
 suggests that an infinite subset of diagrams must be resummed to get a finite conductance \cite{div}. One way to solve this problem is to phenomenologically describe the impurity spin by the Bloch equations \cite{Wolfle,Garst05}
\be		
\hspace{-.079cm}\partial_t\langle S_a\rangle=[\vec h(t)\times\langle\vec S\rangle]_a-\frac{\langle S_a\rangle-\tilde\chi_0 h_a (t)}{\tau_a}, \quad a=x,y,z.\label{eqBloch}
\ee
Here $\vec h(t)=(-t_C,h_y(t),0)$ where $h_y(t)$ is an infinitesimal time-dependent $y$-component of the magnetic field, introduced to obtain $\chi_{yy}$.
We expect Eq.\,\pref{eqBloch} to hold as the equation of motion for the averaged impurity spin in a theory where the quasiparticles are formally integrated out. Here, $\tilde\chi_0\approx-1/4T$ is the static susceptibility $\langle{\vec S}\rangle_0=\tilde\chi_0\vec h$, in the presence of the static field $-t_C$. 
$\langle S_a\rangle(t)$ rotates around the external magnetic field and relaxes towards it within time-scale $\tau_a$ because of its coupling to the quasiparticles. 
Therefore, using the definition $\chi_{yy}(\omega)\equiv \braket{S_y}_\omega/h_y(\omega)$
for the imaginary part of the susceptibility we obtain
\be
\chi''_{yy}(\omega\ra 0)=\frac{\omega\tilde\chi_0 \tau_y}{1+t^2_C \tau_y\tau_z},
\ee
To obtain the conductance, we need $\tau_{y,z}$. The main `Bloch equation' assumption, justifiable at $T\gg T_K$, $t_C$, is to neglect the frequency-dependence of these rates, thus obtaining them from a large frequency limit of our perturbative result using ${\tilde\chi_0}/{\tau_z}=\lim_{\omega\ra\infty} \omega\chi''_{zz}(\omega)\propto \lambda_\perp^2(T/D)^{2\nu-2}$ and ${\tilde\chi_0}/{\tau_y}=\lim_{\omega\ra\infty} \omega\chi''_{yy}(\omega)=\gamma_z\lambda_z^2+\gamma_\perp\lambda_\perp^2(T/D)^{2\nu-2}$. 
So at high temperatures
\be
G_{\nu=1}\propto\frac{t_C^2}{T^2(\lambda_\perp^2+\lambda_z^2)}, \qquad G_{\nu=1/3}\propto\frac{t_C^2}{\lambda_\perp^2T^{2/3}D^{4/3}}.
\ee
(We show explicitly that this result can be obtained, at high $T$, from a resummation of Feynman diagrams 
in the special case $\lambda_y=\lambda_z=0$ in the Supplementary Material.)
More correctly, $\lambda_\perp$, $\lambda_z$ should be replaced by the renormalized quantities at energy scale $T$, but this is an unimportant 
correction assuming $T\gg T_K$. 

\begin{figure}[htp]
\includegraphics[width=1\linewidth]{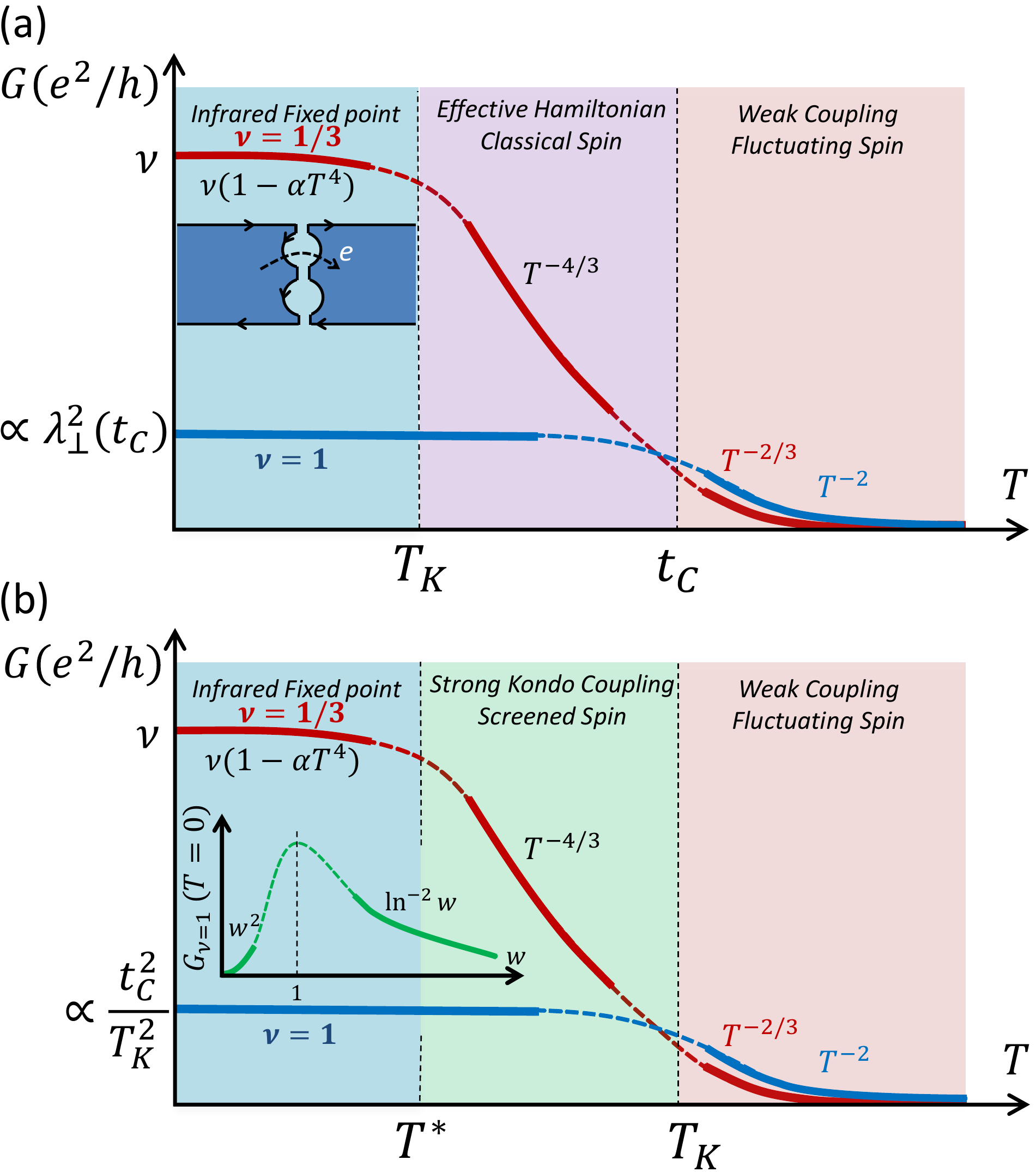}
\caption{\raggedright\small (color online) Inter-edge  conductance vs. temperature for $\nu=1$ and $\nu=1/3$ quantum Hall states at (a) $t_C\gg T_K$ and (b) $t_C\ll T_K$. The dashed lines are interpolations which we expect to be qualitatively correct for the crossover regimes. Whereas the 
$\nu=1$ case exhibits a crossover only at larger crossover scale ${\rm max}(t_C,T_K)$, the $\nu=1/3$ case has an additional crossover at $T_K$ and $T^*$ for the case (a) and (b), respectively. Insets: (a) Schematic of the stable infrared fixed point in the case of $\nu=1/3$. The leading irrelevant processes correspond to electron tunnelling through vacuum. (b) The Kondo fixed point conductance in the case of $\nu=1$ has a non-monotonous dependence on $w\equiv t_C/T_K$, with a peak of ${\cal O}(1)$ at $w\sim 1$.}
\label{fig:fig2}
\end{figure}

\emph{{$T,T_K\ll t_C$} -} In this regime, the impurity spin becomes a classical field pointing in the direction of the instantaneous field, $\vec h(t)$ (see Supplementary Material Sec. V) so we may approximate:
\be H\approx H_0+(1/2)\lambda_\perp [ {\cal J}_x(0)+(h_y(t)/t_C){\cal J}_y(0)].\ee
This corresponds to a direct tunnelling term between edges: $H_T=(1/4)\lambda_\perp [e^{ih_y(t)/t_C}\psi^\dagger_{qp,L}(0)\psi_{qp,R}(0)+h.c.]$. 
For $\nu =1$ this is a simple non-interacting tunnelling model giving a conductance $G\propto \lambda_\perp^2$. [More accurately, $\lambda_\perp$ 
should be replaced by the renormalized coupling $\lambda_\perp (t_C)$ but this is again unimportant for $T_K\ll t_C$.]  For the fractional quantum Hall case 
the behaviour is much different \cite{Moon93} since this direct tunnelling interaction is relevant and $\lambda_\perp (T)=(t_C/T)^{2/3}\lambda_\perp(t_C)$. 
Therefore the conductance starts to grow as $G\propto T^{-4/3}$.  It starts to level off at $T$ of order $T_K$ eventually saturating at $\nu$, corresponding 
to perfect transmission through the double antidots. The nature of this zero temperature infrared fixed point can be straightforwardly understood from 
bosonization. The relevant tunnelling term, $\propto -\lambda_\perp \cos[\sqrt{2}\varphi_s(0)]$, pins
 $\varphi_s(0)$ at $0$.  To understand the physical implications of this boundary condition note that 
while the charge boson remains continuous at $x=0$ at both high $T$ and low $T$ fixed points, 
implying $\varphi_R(0^+)-\varphi_L(0^-)=\varphi_R(0^-)-\varphi_L(0^+)$, the high $T$ and low $T$ 
boundary conditions on the spin boson imply $\varphi_R(0^+)+\varphi_L(0^-)=\pm [\varphi_R(0^-)+\varphi_L(0^+)]$, respectively. Together, these boundary conditions merely imply continuity of $\varphi_{R/L}$ at the 
origin at high $T$ but imply $\varphi_R(0^\pm)=-\varphi_L(0^\pm)$ at low $T$, corresponding to a 
breaking of the system into $x<0$ and $x>0$ parts, perfect transmission through the double 
dots and perfect backscattering. 
The leading low $T$ reduction of the 
conductance is conveniently calculated by considering the small horizontal current between the 
 nearly disconnected $x<0$ and $x>0$ parts of the system [inset of Fig.\,\ref{fig:fig2}(a)]. This involves {\it electrons} tunnelling through vacuum (as opposed to quasiparticle tunnelling through incompressible liquid) between $x<0$ and $x>0$ sides and corresponds to a term in 
the effective Hamiltonian 
$\propto \cos [(\varphi_R(0^+)-\varphi_R(0^-))/\nu]$, of RG scaling dimension $1/\nu$. Thus the 
horizontal conductance, {\it past} the double dots is 
$G_h\propto T^{2/\nu -2}=T^4$. By current conservation, we expect the vertical 
conductance {\it through} the antidots to behave as $G\to \nu-\alpha (T/T_K)^4$, 
for a dimensionless constant of ${\cal O}(1)$, $\alpha$. The behaviour of the conductance when $T_K\ll t_C$ for various temperature ranges and $\nu =1$ and $\nu=1/3$, 
is plotted in Fig.\,\ref{fig:fig2}(a). The zero temperature conductance, as well as the exponens for $T\ll t_C$ agree with previous numerical results \cite{Moon93}.

\emph{{Strong Kondo coupling fixed point, $t_C\ll T\ll T_K$} - }
In this parameter regime, the Kondo  coupling constants $\lambda_\perp$ and $\lambda_z$ renormalize to large values but the interdot tunnelling, $t_C$,
 may be treated as a small perturbation. The impurity spin is then screened by the quasiparticles and, for $\nu =1$, we may apply Fermi liquid theory. 
 The impurity spin, $S^x$ appearing in the interdot tunnelling Hamiltonian can then be represented by $(v/T_K)\psi^\dagger_{qp}(0)\sigma^x\psi_{qp}(0)$, 
 the lowest dimension operator with the correct SU(2) spin transformation properties \cite{Eggert92,IanKondo}. The factor of $v/T_K$ can be inserted by dimensional analysis, 
 recognizing that $T_K$ is the characteristic energy scale, or reduced bandwidth at this fixed point. The corresponding Hamiltonian is non-interacting, 
 with this tunnelling term being marginal under the renormalization group.  This leads to the familiar Shiba formula \cite{Shiba} giving 
 $G\propto (t_C/T_K)^2$.  Similar reasoning may be applied to the $\nu =1/3$ case but now the effective interaction $\propto \psi^\dagger_{qp}(0)\sigma^x\psi_{qp}(0)$
 is relevant, with dimension $1/3$. Thus calculating the conductance to lowest order in $t_C$ gives $G\propto (t_C/T_K)^2(T_K/T)^{2(1-\nu )}\propto T^{-4/3}$. 
 This diverges at low $T$ signalling the breakdown of perturbation theory in $t_C$.
 
 \emph{{$T\ll t_C\ll T_K$ regime- }}
  For $\nu =1$ there is no significant change in behaviour as $T$ is lowered to zero below $t_C$, with the conductance being approximately constant. 
 On the other hand, for $\nu =1/3$, the growth of the interdot tunnelling term under renormalization signals the crossover to the same fixed point 
 discussed above for $T,T_K\ll t_C$, corresponding to perfect transmission through the antidots.  Renormalized interdot tunnelling becomes strong 
 at the scale $T^*\propto t_C^{3/2}/T_K^{1/2}$ and below this scale the conductance should again crossover to 
 $\nu -\alpha (T/T^*)^4$ behaviour.  The behaviour of the conductance when $t_C\ll T_K$ for various temperature ranges 
is plotted in Fig.\,\ref{fig:fig2}(b).  

Note that the conductance versus temperature looks rather similar in the two cases $T_K\ll t_C$ and $t_C\ll T_K$. One essential 
difference is the crossover temperature scales. For $\nu =1$, there is only one crossover which occurs at the larger of $t_C$ and $T_K$. 
For $\nu =1/3$ there are 2 crossover scales: $t_C$ and $T_K$ for $t_C\gg T_K$, but $T_K$ and $t_C^{3/2}/T_K^{1/2}$ for $t_C\ll T_K$. It is also 
interesting to note that, for $\nu =1$ and $\lambda_\perp =\lambda_z$, the $T=0$ conductance is $\propto \lambda_\perp (t_C)^2=1/\ln^2 (t_C/T_K)$ 
for $T_K\ll t_C$ but $\propto (t_C/T_K)^2$ for $t_C\ll T_K$. $G(0)$ decreases as $t_C/T_K$ becomes large or small, going through 
a peak of ${\cal O}(1)$ at $T_K$ of order $t_C$ [see inset of Fig.\,\ref{fig:fig2}(b)]. 

\emph{{Conclusion} -}
We have mapped  the conductance through two antidots in $\nu=1$ integer and $\nu=1/3$ fractional quantum Hall systems onto the susceptibility of a Kondo impurity in a Luttinger liquid, analyzed the fixed points and calculated the conductance in all asymptotic regimes. Calculations of noise, and extension to more exotic filling factors $\nu=5/2$ and $12/5$ with non-abelian statistics are left as future extensions of these results. 

It is a pleasure to thank J.~Folk for bringing up the topic. This research was supported in part by 
NSERC (YK, IA) and CIfAR (IA). PS would like to thank the kind hospitality of the Physics department in UBC where this work was initiated and the french ANR contract Dymesys (ANR 2011-IS04-
001).

\appendix
\section*{Supplementary Material}

In this supplementary material we provide details and proofs of some results presented in the paper. Appendix\,\ref{sec:glueing} discusses the weak coupling and strong Kondo coupling fixed points and the corresponding glueing conditions. Appendix\,\ref{sec:antidots} contains some discussion about the antidots and experimental considerations. Appendix\,\ref{sec:bloch} provides a calculation of the susceptibility using the semiclassical Bloch equation. Appendix\,\ref{sec:pert} provides the result of perturbation theory to second order in Kondo coupling but exact in $t_C$. In Appendix\,\ref{sec:eff} we provide a detailed discussion of the effective Hamiltonian derivation in the case of $T,T_K\ll t_C$. Finally, Appendix\,\ref{sec:xray} contains an exact solution of the case $\lambda_y=\lambda_z=0$, using techniques developed to study the X-ray edge singularity. We show that it is possible to derive the Bloch equation as the high temperature ($T\gg t_C$) formula for the susceptibility. The exact result in this special case also demonstrates the breakdown of perturbation theory at high temperatures.

\section{Gluing conditions at fixed points}\label{sec:glueing}
\subsection{Folding transformation}
For a discussion of the fixed points, it is convenient to fold the chiral bosons, according to
\be
\phi_{s,c}\equiv\frac{\varphi_{s,c}(x)+\varphi_{s,c}(-x)}{\sqrt{2}}
\quad\theta_{s,c}(x)\equiv\frac{\varphi_{s,c}(x)-\varphi_{s,c}(-x)}{\sqrt{2}},\nonumber
\ee
for $x>0$, in terms of which the Hamiltonian becomes
\bea
H&=&H_0+\frac{J_{\perp}}{2}(S^+e^{i\phi_s(0)}+h.c.)
\nonumber\\&&\qquad
+\frac{J_z}{4\pi\nu}S^z\partial_x\theta_s(0)
-t_CS^x+h_zS^z\label{eqHam}
\eea
where we also allowed for a detuning $h_z\equiv\eps_R-\eps_L$ between the two antidots. Here $H_0=\intoinf{\frac{vdx}{4\pi\nu}[(\partial_x\phi_s)^2+(\partial_x\theta_s)^2+(\partial_x\phi_c)^2+(\partial_x\theta_c)^2]}$ and the non-chiral bosons obey $[\phi_s(x),\theta_s(y)]=2\pi i\nu \Theta(x-y)$ where $\Theta(x)$ is the Heaviside step function.
\subsection{Weak coupling}
The only boundary condition at the weak coupling fixed point is that $\varphi_{s,c}(x)$ are continuous at $x=0$, i.e., $\theta_{s,c}(0)=0$. Using the definition
$\com{\int{dx\rho_{R/L}(x)}}{\psi_{qp,R/L}(y)}=-\psi_{qp,R/L}(y)$ we obtain the density operator for both bosons to be $\rho_{R/L}(x)=\frac{1}{2\pi\nu}{\partial_x\varphi_{R/L}}$. From this, assuming a finite length $L$ with periodic boundary condition, the mode expansions are
\bea
\varphi_{R/L}(x)&=&\frac{2\pi\nu}{L} N_{R/L}x+\varphi_{R/L, 0}\nonumber\\
&&\hspace{-1.5cm}+\sum_{n=1}^{\infty}\sqrt{\frac{2\pi\nu}{L k_n}}(\hat a\dn_{R/L, n}e^{\pm ik_nx}+\hat a\dg_{R/L, n}e^{\mp ik_nx})e^{-\frac{k_na}{2}},\qquad\label{eqmode} \eea
Here $k_n={2\pi n}/{L}$, and the harmonic bosons and the zero mode obey the standard commutation relation $[a\dn_{R/L, n},a\dg_{R/L, m}]=\delta_{nm}$ and $[\varphi_{R/L,0},N_{R/L}]=i$. This leads to the mode expansion of the charge/spin bosons
\bea
\varphi_{c/s}(x)&=&\frac{2\pi\nu}{\sqrt{2}L}(N_R\pm N_L)x+\frac{1}{\sqrt{2}}(\varphi_{R0}\mp \varphi_{L0})+\cdots\qquad\label{eqdefboson}
\eea
Defining total bulk charge $Q=N_R+N_L$ and spin $2s_z=N_R-N_L$, for $Q$ even, $s_z$ has to be integer, while for $Q$ odd it must be half-integer. So, the glueing condition for $(Q,s_z)$ at weak coupling fixed point is \cite{IanKondo}
\be
(Q,s_z)=(\textit{even,integer})\oplus(\textit{odd,half-integer}).
\ee
\subsection{Kondo fixed point}
By  power counting, the $J_z$ term is marginal but the $J_\perp$ terms is relevant. In order to account for this, we 
define dimensionless couplings
$\lambda_z=2\pi J_z/v$ and $\lambda_\perp=J_\perp D^{\nu-1}$ and will frequently switch between $J_{z,\perp}$ and $\lambda_{z,\perp}$ notations in the following. These couplings grow as the bandwidth is reduced \cite{PKA} and the system flows to the Kondo fixed point.
Although $J_z$ is naively marginal, because of the coupling to $S^z$, it controls the scaling dimension of the $J_{\perp}$ term. This can be seen if we apply a unitary transformation 
\cite{Emery92,Ingersent95} $H\ra V_\mu\dg H V_\mu$ with $V_\mu=\exp[i\mu S^z\phi_s(0)]$ which gives
\bea
H\ra H_0&+&\frac{J_{\perp}}{2}(S^+e^{i(1-\mu)\phi_s(0)}+h.c.)\nonumber\\
&&+\Big(\frac{J_z}{4\pi\nu}-\mu v\Big)S^z\partial_x\theta_s(0)+h_zS^z\nonumber
\\&&-t_C[S^x\cos\mu\phi_s(0)+
S^y\sin\mu\phi_s(0)]\label{eqEmery}
\eea
and changes the dimension of $J_\perp$ from $1-\nu$ to $1-\nu(1-\mu)^2$.
In order to understand the strong coupling Kondo fixed point, it is convenient to either tune this dimension to zero (the so-called Toulouse \cite{Toulouse69} point) so that it could be refermionized \cite{Ingersent95} or to 1 (the so-called decoupling poing) so that it becomes a boundary magnetic field. We use the latter approach, which happens at $\mu=1$ and the transverse Kondo coupling becomes $J_\perp S^x$. In the case of $t_C=h_z=0$, and if $\lambda_z=4\pi\nu$, the Kondo coupling reduces to a Zeeman field on the isolated (but dressed) impurity spin which projects to the ground state of $S^x$ at low energies. The Kondo temperature is $\sim J_\perp$ in this highly anisotropic Kondo model \cite{Laflorencie08}. It is easy to check that $\varphi_s(x)$ develops a discontinuity at $x=0$
\be
\tilde\varphi_s(x)=V_\mu\dg\varphi_s(x)V\dn_\mu=\varphi_s(x)-\pi\nu\mu S^z\sgn(x)
\ee
or equivalently, in the folded basis, the new boundary condition corresponds to $\tilde\phi_s(0)=\phi_s(0)$ and a new pinning of $\tilde\theta_s(0)=\mp\pi\nu$. The pinning is dynamically switching between these two values as in the instanton-gas representation of the Kondo problem \cite{AndersonYuval}.
The charge boson is unchanged by the unitary transformation followed by the projection and the new gluing condition at the strong Kondo coupling fixed point \cite{IanKondo} is
\be
(Q,s_z)=(\textit{even,half-integer})\oplus(\textit{odd,integer}).
\ee
This change in the glueing condition, implies that a spin boson has decoupled from the edge states to screen the impurity spin. We have used the decoupling point to discuss the Kondo fixed point, and this requires tuning $J_z$ to the large value of $4\pi v\nu$ which is not physical, as the bare Kondo coupling is usually assumed to be small. However, it is expected that other values of $J_z$ would have a similar qualitative behaviour.

\section{Antidots}\label{sec:antidots}
We can find the spectrum of the antidot by inserting the mode expansion \pref{eqmode} into the free Hamiltonian
\be
H=\frac{v}{2\pi\nu}\int_0^L{dx(\partial_x\varphi)^2}=\sum_{k>0}vka\dg_k a\dn_k+E_Cn^2,\label{eq11}
\ee
where $E_C={2\pi v\nu}/{L}$ acts like the ``charging energy'' of the antidot and $L$ is its circumference.
We see that the number of quasiparticles $n$ is a good quantum number $\hat n\ket{n}_0=n\ket{n}_0$. These two sectors are coupled to each other because $L=2\pi R$ and $R=\sqrt{2N}\ell_B$ in terms of magnetic length $\ell_B=\sqrt{\hbar/eB}$, but we can assume that the antidots are large enough so that the radius of $\ket{n}_0$ and $\ket{n+1}_0$ are effectively the same [\onlinecite{Averin07}] and assume that charge and neutral sectors decouple. This corresponds to the constant interaction model in quantum dots \cite{Pustilnik}. 
For the antidots, we are interested in a regime where bosonic excitation energy $\delta\eps=2\pi v/L$ is much larger than $k_BT$ and we can assume that harmonic part of the field is in its ground state $a_k\ket{0}_n=0$. Note that $\ket{n}_0$ and $\ket{\varphi}_n$ are analogous to the number of charges and the excited states of normal quantum dots.

We also need to take into account the Aharonov-Bohm contribution of the magnetic flux going through the antidots. The number of quasiholes $n_R$ in antidot $R$ is such that it is equal to the number of flux quanta going through $\phi_R/\phi_0=n_R$ where $\phi_0=h/e$. This is another way of stating that $R_R=\sqrt{2n_R}\ell_B$, at the ground state. These numbers change as we change $\phi_R$. This is done by replacing $E_Cn_R^2$ in Eq.\,\pref{eq11} by $E_C(n_R-\phi_R/\phi_0)^2$ with $\varphi_0\equiv h/e$.
We are interested in a regime where two states in the same antidot with $n_R$ and $n_R+1$ quasiparticles become degenerate. This is possible for $\phi_R/\phi_0=2m_R+1$ where $m_R$ is an integer. Also we need similar degeneracy to be valid for the second antidot $\phi_L/\phi_0=2m_L+1$. To have both of these at the same magnetic field, we obviously need some fine tuning of the area of at least one of the antidots. We assume that this is possible by tuning the voltage applied to the gates that defined the antidots at the first place, or by a combination of voltages applied to the outer edges. To capture deviation from this ideal case one can add a term $h_zS^z$ to the Hamiltonian, but we assumed a perfect tuning in the paper.

If the temperature is low enough ($ T\ll v_F/L\ll E_F$) so that the bosonic modes of the antidots are not excited, they effectively behave as hardcore fermions [\onlinecite{Averin07}]. To see this, following \cite{Affleck14} we assume that $N$ and $N+1$ states of the antidot are degenerate and denoting them by $\ket{0}$ and $\ket{1}$, it can be seen that due to the commutation relation $[\varphi_0,N]=i$, the operators $s^{\pm}\propto e^{\pm i\varphi_0}$ are raising and lowering operators of the "spin" made of $\ket{0}$, $\ket{1}$. From $[\varphi_0,N]=i$, it follows
\be
[N,e^{\pm i\varphi_0}]=\pm e^{\pm i\varphi_0} \andd Ne^{\pm i\varphi_0}=e^{\pm i\varphi_0}(N\pm 1)\nonumber
\ee
These can be combined with the bosonization Klein factors $\Gamma_{L,R}$ to represent the creation and 
annihilation operators for the additional fermion on the dot.
\be
d\dg_L\equiv\Gamma_L e^{i\varphi_{L0}} \andd d\dg_R\equiv\Gamma_R e^{i\varphi_{R0}}
\ee

\section{Bloch equation - Non-zero $t_C$\label{sec:bloch}}
Considering that $\vec h=(-t_C,h_y,0)$ and $\lambda_y\neq\lambda_z$, there is no spin symmetry present and we have to allow for different relaxation rates along each direction. Therefore, we can write the Bloch equations [Eq.(4) of the paper] as
\bea
\partial_t\braket{S_x}&=&h_y\braket{S_z}-\frac{\braket{S_x}-\braket{S_x^0}}{\tau_x}, \\
\partial_t\braket{S_y}&=&t_C\braket{S_z}-\frac{\braket{S_y}-\braket{S_y^0}}{\tau_y}, \\
\partial_t\braket{S_z}&=&-t_C\braket{S_y}-h_y\braket{S_x}-\frac{\braket{S_z}-\braket{S_z^0}}{\tau_z}.
\eea
where $\braket{S_a^0}$ are the components of the steady state magnetization.
To find the steady state magnetizations, we do a rotation ($\tan\vartheta=-h_y/t_C$)
\be
O(\vartheta)=\mat{\cos\vartheta & -\sin\vartheta \\ \sin\vartheta & \cos\vartheta}
\ee
on
\be
\mat{\tilde h_x & 0}=\mat{-t_C & h_y}
O(\vartheta)
, \qquad \mat{S_x \\S_y}=O(\vartheta)\mat{\tilde S_x\\  \tilde S_y}\nonumber
\ee
to obtain $\tilde h_x=\cos\vartheta (-t_C+h_y^2/t_C)=-t_C+{\cal O}(h_y^2)$. The Hamiltonian is diagonal in this `tilde' basis and we find
\bea
\braket{S_x^0}&=&\cos\vartheta\braket{\tilde S_x^0}=-\frac{1}{2}\tanh\frac{\tilde h_x\beta}{2}\cos\vartheta,\\
\braket{S_y^0}&=&-\frac{1}{2}\tanh\frac{\tilde h_x\beta}{2}\sin\vartheta, \qquad
\braket{S_z^0}=0
\eea
Since eventually we are interested in $\chi_{yy}=d\braket{S}_y/dh_y\vert_{h_y=0}$ we can drop ${\cal O}(h_y^2)$ and the above results simplify to
\bea
\braket{S_x^0}&\approx& \frac{1}{2}\tanh\frac{t_C\beta}{2}, \\
 \braket{S_y^0}&\approx& -\frac{1}{2}\frac{h_y}{t_C}\tanh\frac{t_C\beta}{2}, \qquad \braket{S_z^0}=0
\eea
For $T\gg t_C$ where we expect the Bloch equation approach to be valid, $\tanh(t_C\beta/2)\approx t_C\beta/2$ and we get the linear response result $\braket{S_y^0}\approx\chi_0h_y$, but for $T\ll t_C$, we have $\tanh(t_C\beta/2)\approx 1$. More generally, we can define an effective static susceptibility given by $\langle{\vec S^0}\rangle\approx\tilde\chi_0\vec h$, where
\be
\tilde\chi_0(T)\equiv\frac{-1}{2t_C}\tanh\frac{t_C\beta}{2}=\chi_0\Big(T\ra \frac{t_C}{2\tanh\frac{t_C\beta}{2}}\Big)
\ee
Fourier transforming, we obtain 
\bean
(1-i\omega \tau_x)\braket{S_x}_\omega&=&-2\pi\delta(\omega)\chi^{\rm eff}_0t_C+\tau_x
h_y(\omega)*\braket{S_z}_\omega\\
(1-i\omega \tau_y)\braket{S_y}_\omega&=&\chi^{\rm eff}_0h_y(\omega)+t_C\tau_y\braket{S_z}_{\omega}\\
(1-i\omega \tau_z)\braket{S_z}_\omega&=&-t_CT^z_2\braket{S_y}_\omega-\tau_z
h_y(\omega)*\braket{S_x}_\omega
\eean
These are easily generalized to the memory-full case, by allowing a frequency-dependence for $T_a(\omega)$.
These set of equations are difficult to solve. One approximation that greatly simplifies this, is to ignore the fluctuations of the spin along the external field, $x$ direction. This amounts to dropping the second (convolution) term on the right hand side of the first equation and makes sense because we expect the second term to be ${\cal O}(h_y^2)$. Then everything simplifies: We get $\braket{S_x}_\omega\approx -2\pi\delta(\omega)\tilde\chi_0t_C$, i.e. constant in time. Thus the convolution in the last line also simplifies and we obtain
\be
(1-i\omega \tau_z)\braket{S_z}_\omega\approx -t_C\tau_z\Big[\braket{S_y}_\omega-\tilde\chi_0h_y(\omega)\Big],
\ee
from which we get 
\be
\chi_{yy}(\omega)\equiv\lim_{h_y\ra 0}\frac{\braket{S_y}_\omega}{h_y(\omega)}=\frac{\tilde\chi_0(1-i\omega \tau_z+t_C^2\tau_z\tau_y)}{(1-i\omega \tau_z)(1-i\omega \tau_y)+t_C^2\tau_z\tau_y}\label{eqWolfle}
\ee
with the imaginary part
\be
\chi''_{yy}(\omega)=\frac{\omega\tilde\chi_0 \tau_y[1+t_C^2 \tau_z\tau_y+\omega^2 \tau_z^2]}{[1+(t_C^2-\omega^2) \tau_z\tau_y]^2+\omega^2 (\tau_y+\tau_z)^2}
\ee
Let us look at this formula, in various limits. Without Kondo coupling $ T_{x,y,z}\ra\infty$, and we get
\be
\chi_{yy}(\omega)\ra-\frac{t_C^2\tilde\chi_0}{\omega^2-t_C^2}=\frac{t_C}{2}\frac{\tanh(\beta t_C/2)}{\omega^2-t_C^2}
\ee
For $t_C\sqrt{\tau_z\tau_y}\ll 1$, we basically get the simple result $\chi_{yy}(\omega)=\tilde\chi_0/(1-i\omega \tau_y)$ that we would get if we had neglected $t_C$ from the beginning. Generally, we see that 
\be
\hspace{-.1cm}\chi''_{yy}(\omega\ra 0)=\frac{\omega\tilde\chi_0 \tau_y}{1+t^2_C \tau_y\tau_z},\qquad G=\frac{-8\pi t_C^2 \tilde\chi_0 \tau_y}{1+t^2_C \tau_y\tau_z}\label{eqfullbloch}
\ee
But for $\omega\ra\infty$
\be
\hspace{-.1cm}\chi''_{yy}(\omega\ra\infty)=\frac{\tilde\chi_0}{\omega \tau_y}, \quad\ra\quad \frac{\tilde\chi_0}{\tau_y}=\lim_{\omega\ra\infty} \omega\chi''_{yy}(\omega)
\ee
To obtain the conductance, we also need $\tau_z$ which can be obtained using
\be
\frac{\tilde\chi_0}{\tau_z}=\lim_{\omega\ra\infty} \omega\chi''_{zz}(\omega)
\ee
or from $\tau_y$ with a rotation along $S_x$, i.e. by interchanging $\lambda_y\lr\lambda_z$.

\section{Susceptibility to order ${\cal O}(\lambda^2)$ but exact in $t_C$\label{sec:pert}}
In this section we provide the result of perturbative calculations of the imaginary part of the susceptibility to second order in Kondo coupling but exact in inter antidot tunneling $t_C$. The goal of this section is to demonstrate that once a finite $t_C$ is included the infrared divergence of the perturbation theory is cut off. Using equation of motion techniques, it can be shown that the correlation functions to second order in Kondo coupling are
\bea
\chi{''}_{yy}^{(zz)}=\frac{\lambda_z^2}{16}\frac{\omega^2}{(\omega^2-t_C^2)^2}\im{\Pi^R_{zz}(\omega)}\\
\chi{''}_{yy}^{(zz)}=\frac{\lambda_y^2}{16}\frac{t_C^2}{(\omega^2-t_C^2)^2}\im{\Pi^R_{yy}(\omega)}
\eea
Here $\Pi_{aa}^R(\omega)\sim\braket{{\cal J}_a{\cal J}_a}_\omega$ are retarded correlation functions of the current operators,
\bea
&&\Pi_{yy}^R(\omega,\nu=1)=\Pi^R_{zz}(\omega,\nu=1), \qquad
\Pi^R_{zz}(\omega,\nu)
=\frac{-i\omega}{8\pi\nu v^2}\nonumber\\
&&\Pi_{yy}^R(\omega,\nu<1/2)=-\Big(\frac{2\pi}{\beta}\Big)^{2\nu-1}\hspace{-.25cm}\sin(\pi\nu)B\Big(\nu-\frac{i\omega\beta}{2\pi},1-2\nu\Big)\nonumber
\eea
where $B(x,y)$ is the beta function. The $\lambda_x^2$ contribution has a more complicated form
\bea
\chi_{yy}''(\omega<t_C)&=&-g\Big\{\frac{\pi}{2}\frac{\omega}{\omega^2-t_C^2}+\nonumber\\
&&\hspace{-1.1cm}+\frac{1}{2}\tanh(\beta t_C/2)\Big[\frac{2\pi}{\beta}\frac{t_C}{\omega^2-t_C^2}+H_{\nu}(\omega)\Big]\Big\}\label{eqpertxtc}
\eea
where he function $H_{\nu}(\omega)$ for $\nu=1$ is
\be
H_{\nu=1}(\omega)\equiv\im{\frac{\psi\Big(1-\frac{i\beta(\omega+t_C)}{2\pi}\Big)}{\omega+t_C}-\frac{\psi\Big(1+\frac{i\beta(t_C-\omega)}{2\pi}\Big)}{\omega-t_C}}\nonumber
\ee
in terms of digamma function $\psi(z)$. Note that in the limit of $t_C\ra 0$, these results reduce to Eq.\,\pref{eqchip} of the paper.
\section{Effective Hamiltonian for $T,T_K\ll t_C$\label{sec:eff}}
Temporarily igoring the infinitesimal time-dependent part of the field, $h_y(t)$, introduced to calculate the dynamical susceptibility, 
$\chi_{yy}''$, it is clear that in this regime we may replace $S^x$ by $1/2$ (and $S^{y,z}$ by zero) since the impurity spin is polarized by the strong field. 
Now consider the effect of $h_y(t)$. We again wish to integrate out the impurity spin to obtain an effective Hamiltonian for the 
quasiparticles. It is now not appropriate to consider ant relaxation terms in the Bloch equations, since such terms arise from the integrating out the quasiparticles instead. So, let's consider the solutions of the 
simple spin torque equation
\be \partial_t\vec  S=\vec h(t)\times \vec S(t)\ee
with $\vec h(t)\equiv (-t_C,h_y(t),0)$, $h_y(t)=\epsilon t_C\cos \omega t$, taking the limit where both $\epsilon \to 0$ and $\omega \to 0$. 
We also assume that the oscillating component of the field is turned on slowly in the infinite past.  Thus we write:
\be \vec S(t)=(1/2)(1,0,0)-\vec S'.\ee
We will see that $\vec S'$ is $O(\epsilon )$. 
Working to first order in $\epsilon$, 
\be \vec h\times S/t_C\approx  -(1/2)(0,0,\epsilon \cos \omega t)+(0,-S^z{'},S^y{'})
\ee
Thus
\bea {1\over t_C}{dS^z{'}\over dt}&=&S^y{'}-{\epsilon \over 2}\cos \omega t\\
{1\over t_C}{dS^y{'}\over dt}&=&-S^z{'}.
\eea
Thus
\be {1\over t_C}{d^2S^z{'}\over dt^2}=-t_CS^z{'}+{1\over 2}\epsilon \omega \sin \omega t.\ee
with solution
\be S^z{'}\approx {\epsilon t_C\omega \over 2(t_C^2-\omega^2)}\sin \omega t\approx {\epsilon \omega \over 2t_C}\sin \omega t
\ee
Thus
\be {dS^y{'}\over dt}\approx -{\epsilon \omega \over 2}\sin \omega t\ee
with solution
\be S^y{'}={\epsilon \over 2}\cos \omega t.\ee
In summary
\be \vec S\approx (1/2)(1,-\epsilon \cos \omega t,-\epsilon (\omega /t_C)\sin \omega t).\ee
At small $\omega$ we may drop the last component, giving
\be \vec S\approx -(1/2t_C)\vec h(t).\ee
 Thus we see that, even with no relaxation term in the Bloch equations, purely a precession term, 
the spin tracks the instantaneous time-dependent field, in the limit where the time-dependent term in the field is small 
and slowly varying. 

\section{Exact spin susceptibility when $\lambda_y=\lambda_z=0$}\label{sec:xray}
\subsection{Derivation of Bloch equation result for $t_C\approx 0$}
The fact that Eqs.\,(3) and (6) of the paper contain a simple summation of $\lambda_\perp^2$ and $\lambda_z^2$ contributions to lowest order, suggests that the same IR divergence and the necessity to use Bloch equation occurs even if only one Kondo coupling, say $\lambda_z\neq 0$, is non-zero. In that limit and assuming we could neglect $t_C\approx 0$ at $T\gg t_C$, it is possible to map our Kondo problem to the X-ray absorption problem and find a non-perturbative formula for the susceptibility. 


With only $\lambda_z$ nonzero, the Hamiltonian is $H=H_0+\lambda_zS^z{\cal J}_z(0)$ and the Kondo interaction is just a boundary magnetic field, depending on the spin state of the impurity. We are interested in the dynamic susceptibility $\braket{S^yS^y}_\omega$, defined after Eq.\,(2) of the paper. But $S^y$ itself, is not present in the Hamiltonian and its influence is just suddenly switching the sign of the boundary magnetic field, via $S^y\ket{\ua}=i\ket{\da}$ and $S^y\ket{\da}=-i\ket{\ua}$. Before this switching, the spin-up fermions see a phase shift and spin-down fermions another and these two phase shifts are suddenly switched. The ground states before and after switching are orthogonal to each other in the thermodynamic limit \cite{Mahan} and the transition creates lots of electron-hole pairs, the so-called orthogonality catastrophe. It is more convenient to discuss this in terms of the spin-up/down bosons. Setting the velocity to 1, dropping constants and following \cite{Affleck94} by introducing $\varphi_{\ua/\da}\equiv\varphi_{R/L}$, the Hamiltonian density at $\nu=1$ is
\be
{\cal H}={\cal H}_0+\delta(x)\frac{J_z}{4\pi}S^z(\partial_x\varphi_\ua-\partial_x\varphi_\da),
\ee
where ${\cal H}_0=(\partial_x\varphi_\ua)^2+(\partial_x\varphi_\da)^2$.
Depending on the state of the impurity spin $\ket{\ua}$ or $\ket{\da}$, the Hamiltonian breaks into two sectors ${\cal H}={\cal H}_{+}\ket{\ua}\bra{\ua}+{\cal H}_{-}\ket{\da}\bra{\da}$, where using $\lambda_z=\varrho J_z=J_z/2\pi v$ we get
\be
{\cal H}_\pm = [\partial_x\varphi_\ua\pm \frac{\pi\lambda_z}{2}\delta(x)]^2+[\partial_x\varphi_\da\mp \frac{\pi\lambda_z}{2}\delta(x)]^2,
\ee
up to a constant. Defining $\lambda'\equiv\lambda_z/4$, these two Hamiltonians are related to ${\cal H}_0$ by the Schotte-Schotte unitary transformation \cite{Affleck94}
\be
U=e^{-i\lambda'\varphi_\ua(0)}e^{+i\lambda'\varphi_\da(0)}\label{eq5}
\ee
so that
\be
{\cal H}_+=U\dg{\cal H}_0 U, \qquad
{\cal H}_-=U\dn{\cal H}_0 U\dg,\label{eqA4}
\ee
Writing $S^y$ as $2iS^y=S^+ - S^-$ and applying the unitary evolution operator 
\be
2iS^y(t)=e^{itH_+}S^+ e^{-itH_-}-e^{itH_-}S^- e^{-itH_+}\label{eq7}.
\ee
Here, we used that before/after applying $S^+$, the system has to be in the $-/+$ sectors, respectively. Inserting this into dynamic susceptibility and dropping $S^\pm S^\pm$ terms, we obtain
\bea
\chi^R(t)&\propto&\Theta(t)\Big[\braket{e^{itH_+}e^{-itH_-}S^+S^-}
+\braket{e^{itH_-}e^{-itH_+}S^-S^+}\nonumber\\
&&-\braket{e^{itH_+}e^{-itH_-}S^-S^+}-\braket{e^{itH_-}e^{-itH_+}S^+S^-}\Big]\nonumber
\eea
Using Eqs.\,\pref{eqA4} we can write $e^{\pm itH_+}=U\dg e^{\pm itH_0} U$ and $e^{\pm itH_-}=U e^{\pm itH_0} U\dg$. We have to apply the same procedure as in Eq.\,\pref{eq7} to the Boltzman factors. If the correlation function contains $S^+S^-$, the Boltzman factor becomes $e^{-\beta H}/Z\ra e^{-\beta H_+}/Z=U\dg e^{-\beta H_0} U/2Z_0$ and similar version for the $S^-S^+$ terms. After these substitutions, the spins have done their job and can be simply dropped from the correlation functions and we arrive at
\bea
\chi^R_{zz}(t)&\propto &\Theta(t)\Big[\braket{ e^{iH_0t}U^2e^{-iH_0t}U^{\dagger 2}}+\braket{e^{iH_0t}U^{\dagger 2}e^{-iH_0t}U^2}\nonumber\\
&&\quad-\braket{U^{\dagger 2} e^{iH_0t}U^2e^{-iH_0t}}-\braket{U^2e^{iH_0t}U^{\dagger 2}e^{-iH_0t}}\Big]\nonumber\\
&\propto&\Theta(t)\Big[\braket{U^2(t)U^{\dagger 2}}+\braket{U^{\dagger 2}(t)U^2}
\nonumber\\&&\quad
-\braket{U^{\dagger 2}U^2(t)}-\braket{U^2 U^{\dagger 2}(t)}
\Big].
\eea
The first term $\braket{U^2(t)U^{\dagger 2}}$ at $T=0$ reduces to
\bea
&&\braket{e^{-2i\lambda'\varphi_\ua(t)}e^{2i\lambda'\varphi_\ua(0)}}\braket{e^{2i\lambda'\varphi_\da(t)}e^{-2i\lambda'\varphi_\da(0)}}\nonumber\\
&&=
e^{4i\pi\lambda'^2}\braket{e^{-2i\lambda'[\varphi_\ua(t)-\varphi_\ua(0)]}}\braket{e^{2i\lambda'[\varphi_\da(t)-\varphi_\da(0)]}}\nonumber\\
&&=
\frac{e^{4i\pi\nu\lambda'^2}}{t^{8\nu\lambda'^2}}.
\eea
Doing similar procedure for the other terms and summing up all the terms, at zero temperature we obtain
\be
\chi^R_{zz}(t)=-\frac{\Theta(t)}{2t^{2g}}\sin{\pi g}
,\qquad g\equiv{4\lambda'^2}=\frac{\lambda^2_z}{4}
\ee
This correlation function is a power-law in absence of any bulk interaction, because of the physics of orthogonality catastrophe \cite{Affleck94,Mahan}. Using conformal mapping to a finite-radii cylinder $t\ra \frac{\beta}{\pi}\sinh \frac{\pi t}{\beta}$,
we can bring this retarded function to the finite temperature,
\be
\chi^R(t)=-\Big(\frac{\pi}{\beta}\Big)^{2g}\frac{\Theta(t)\sin\pi g}{2\abs{\sinh\frac{\pi t}{\beta}}^{2g}}\label{eq17}.
\ee
The Fourier transform \cite{Schulz86} gives
\be
\chi^R_{zz}(\omega)=-\Big(\frac{2\pi}{\beta}\Big)^{2g-1}B(g-i\omega\beta/2\pi,1-2g)\frac{\sin\pi g}{2}\nonumber
\ee
We are interested in the limit of small Kondo coupling, $g\ra 0$. Therefore, using properties of the beta function
\be
\chi^R_{zz}(\omega)\approx-\Big(\frac{2\pi}{\beta}\Big)^{-1}B(g-i\omega\beta/2\pi,1)\frac{\pi g}{2}=\frac{\chi_0}{1-i\omega\tau_K}\nonumber
\ee
where $\chi_0=-\beta/4$ and $\beta/\tau_K=2{\pi g}=2\pi{\lambda_z^2}/{4}$. We have dropped $g$ in the second argument of the beta function, but kept it on the first argument. This is what one would obtain assuming $t_C\approx 0$ in Eq.\,\pref{eqWolfle} which relies on the Bloch equation approach. This form was proved using some analytical assumption \cite{Wolfle} or by using the phenomenological Bloch equation \cite{Garst05}, but we provided an exact derivation here, when $\lambda_{x,y}=0$. We expect similar result for $\lambda_x\neq 0$ but $\lambda_{z,y}=0$ by a spin rotation along $y$ direction.

\subsection{Non-zero $t_C$}
When $t_C$ is non-zero, for $\lambda_y\neq0$ or $\lambda_z\neq0$ the Hamiltonian contains non-commuting spin terms and the problem is complicated. However, the special case of only $\lambda_x\neq0$ (but $\lambda_{y,z}=0$) can be still solved exactly using techniques similar to those described above. Note that having $\lambda_x\neq\lambda_y$ is unphysical as a Schrieffer-Wolff transformation would always produce equal transverse Kondo couplings. Nevertheless, this unphysical case can be used as a check on our Bloch equation result. Following the same technique as in previous section, it can be easily shown that
\be
\chi^R_{yy}(t)=\frac{i\Theta(t)}{4[\frac{\beta}{\pi}\sinh(\pi t/\beta)]^{2g}}\Big\{e^{-it_Ct}A-
e^{it_Ct}A^*\Big\},\label{eqtime}
\ee
where
\be
A=-\cos(\pi g)\tanh(\pi b)+i\sin(\pi g).
\ee
\begin{figure}[rtp!]
\includegraphics[width=1\linewidth]{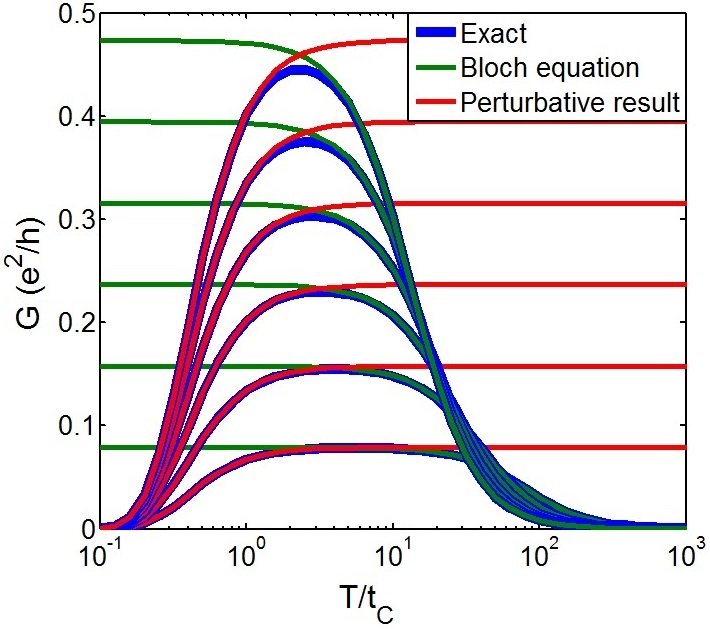}
\caption{\raggedright\small (color online) The special unphysical case of $\lambda_y=\lambda_z=0$, but $\lambda_x\neq 0$ and $t_C\neq 0$. Conductance $G(b,g)$ as a function of $T/t_C$ for various values of $g=\lambda_x^2/4$ on log scale. These values are $g=0.002$, $0.004$, $0.006$, $0.00$8, $0.010$, and $0.012$, corresponding the lowest to highest conductance.
The exact result is compared with perturbation theory to second order in Kondo coupling and exact in $t_C$ (red color) and the Bloch equation result (green).}\label{fig1sup}
\end{figure}
and $b=\beta t_C/2\pi$. 
The Fourier transform of \pref{eqtime} gives
\bea
-4i\chi^R_{yy}(\omega)&=&\Big(\frac{2\pi}{\beta}\Big)^{2g-1}\Big[AB\Big(g-\frac{i\beta(\omega-t_C)}{2\pi},1-2g\Big)\nonumber\\
&&\quad-A^*B\Big(g-\frac{i\beta(\omega+t_C)}{2\pi},1-2g\Big)\Big]
\eea
Using Eq.\,(2) of the paper, the conductance can be written as a closed formula
\bea
G&=&-4\pi b^2{\rm Im}\Big\{A(b,g)B(g+ib,1-2g)\nonumber\\
&&\hspace{2.5cm}\times
[\psi(g+ib)-\psi(1-g+ib)]\Big\}\qquad
\eea
where again $\psi(z)$ is the digamma funciton. Note that the conductance is a function of $b=\beta t_C/2\pi$ and $g=\lambda_x^2/4$ only. Higher values of Kondo coupling squared $g$ correspond to larger conductance. This function is plotted in Fig.\,\ref{fig1sup}, as a function of $(2\pi b)^{-1}=T/t_C$ for various values of $g$, and it is compared with perturbation theory result [Eq.\,\pref{eqpertxtc}] and the Bloch equation result [Eq.\,\pref{eqfullbloch}]. Although the second order perturbation theory (but exact in $t_C$) is sufficient at $T\ll t_C$, it fails in the opposite regime of $T\gg t_C$ as pointed out in the paper. On the other hand, the Bloch equation result provides an accurate estimation of the conductance in this high temperature regime of $T\gg t_C$.


\begin{thebibliography}{50}
\bibitem{fnote1}
In general a potential scattering term of  the form $V\psi^\dagger_{qp,\alpha}(0)\psi_{qp,\alpha}(0)$ is also generated due 
to both the Coulomb interaction
 and to particle-hole asymmetric antidot energies $\epsilon d^\dagger_\alpha d_\alpha$.  We assume $V$ is tuned to zero 
by adjusting gate voltages, as is an effective longitudinal magnetic field term $h_zd^\dagger_\alpha \sigma^z_{\alpha \beta}d_\beta /2$.
\end{thebibliography}
\end{document}